\documentstyle[aps,prl,preprint,floats,epsfig]{revtex}
\def\spwidth{\epsfxsize=5.0in}

\newcommand{\bdg}{{\cal B}(D^{*+}\rightarrow D^+\gamma)}

\newcommand{\Rg}{R_\gamma^+}
\newcommand{\Rp}{R_\pi^+}

\begin{document}
\preprint{\tighten\vbox{\hbox{\hfil CLNS 97/1518}
                        \hbox{\hfil CLEO 97-25}
}}
\draft 

\title{Observation of the Radiative Decay $D^{*+}\rightarrow D^+ \gamma$}

\author{CLEO Collaboration}
\date{\today}

\maketitle
\tighten

\begin{abstract} 
We have observed a signal for the decay $D^{*+} \to D^+ \gamma$ at a
significance of 4 standard deviations.
From the measured branching ratio  
${\cal B}(D^{*+}\rightarrow D^+\gamma)/
      {\cal B}(D^{*+}\rightarrow D^+\pi^{0})= 0.055 \pm 0.014 \pm 0.010 $ 
we find  $\bdg = 0.017 \pm 0.004 \pm 0.003 $,
where the first uncertainty is statistical and the second is systematic.
We also report the highest precision measurements of the remaining $D^{*+}$ 
branching fractions.
\end{abstract}
\newpage

{
\renewcommand{\thefootnote}{\fnsymbol{footnote}}
\begin{center}
J.~Bartelt,$^{1}$ S.~E.~Csorna,$^{1}$ V.~Jain,$^{1,}$%
\footnote{Permanent address: Brookhaven National Laboratory, Upton, NY 11973.}
K.~W.~McLean,$^{1}$ S.~Marka,$^{1}$
R.~Godang,$^{2}$ K.~Kinoshita,$^{2}$ I.~C.~Lai,$^{2}$
P.~Pomianowski,$^{2}$ S.~Schrenk,$^{2}$
% ...
\end{center}
 
\small
\begin{center}
$^{1}${Vanderbilt University, Nashville, Tennessee 37235}\\
$^{2}${Virginia Polytechnic Institute and State University,
Blacksburg, Virginia 24061}\\
% ...
\end{center}
\setcounter{footnote}{0}
}
\newpage

The decays of the excited charmed mesons, $D^{*+}$ and $D^{*0}$, have 
been the subject of extensive 
theoretical~\cite{theory,HHCPT,HHCPT_gB,Lepage} 
as well as experimental~\cite{M1,M2,JADE,HRS,M3,CLEOII,ARGUS} 
investigation.
The decay of the $D^{*0}$ via emission of a $\pi^0$ or a photon has
been observed and its branching ratio well measured~\cite{PDG}.  
While the $D^{*+}$ hadronic decays ($D^{*+}\rightarrow D^+\pi^0$ and
$D^{*+}\rightarrow D^0\pi^+$)~\cite{CC} have been observed and are widely used 
to tag $b$ quark decays, the observation of the $D^{*+}$ radiative decay 
remained problematic.
Both $D^{*}$ mesons decay electromagnetically as the result of a 
spin-flip of either the charm quark or the light quark.
In the case of the $D^{*0}$, the decay amplitudes for these two processes 
interfere constructively.  Combined with the phase space suppression
of the hadronic decay, this interference results in a radiative decay fraction 
which competes with the hadronic decay fraction.
In the case of the $D^{*+}$, the amplitudes for the two spin-flip processes 
interfere destructively.  Also, there is slightly more phase space available 
for the hadronic decay.
These two conditions result in a radiative decay fraction of the $D^{*+}$ 
which, in comparison to the $D^{*0}$, is significantly suppressed relative 
to the hadronic decay fraction.

A great deal of interest in the radiative $D^{*+}$ decay was generated by
an earlier Particle Data Group average of $\bdg=(18\pm4)\%$~\cite{oldPDG};
this value was virtually impossible to reconcile with theory without 
assuming an anomalously large magnetic moment for the charm 
quark~\cite{Lepage}.
Based on 780 pb$^{-1}$ of data, a previous CLEO~II analysis~\cite{CLEOII} 
found an upper limit of $4.2\%$ (90\% C.L.) for this branching fraction,
a result which strongly affected not only the $D^{*+}$ branching fractions
but also many $B$ measurements.
In addition to its importance in measuring $B$ meson decays, a precision
determination of the $D^{*+}$ branching fractions will provide an important 
test of many quark models and other theoretical approaches to
heavy meson decays~\cite{theory}.  For theories built 
around chiral and heavy-quark symmetry (heavy hadron chiral perturbation
theory)~\cite{HHCPT}, this measurement will also provide a strong 
constraint on the two input parameters 
($g$ and $\beta$) allowing model-independent predictions to be made on 
a wide variety of observable quantities~\cite{HHCPT_gB}.

The approach used in this analysis is to search in the 
$\Delta M_\gamma \equiv M(D^+ \gamma)-M(D^+)$~\cite{M_def} and
$\Delta M_\pi \equiv M(D^+ \pi^0)-M(D^+)$
distributions for $D^{*+}$ events using the decay chain 
$D^{*+}\rightarrow D^+(\gamma$ or $\pi^{0}$),
$D^+ \to K^-\pi^+\pi^+$.
The branching ratio 
\begin{equation}
 \Rg \equiv \frac{{\cal B}(D^{*+}\rightarrow D^+\gamma)}
                  {{\cal B}(D^{*+}\rightarrow D^+\pi^{0})}
   =\frac{N(D^+\gamma)}{N(D^+\pi^{0})} \times 
    \frac{\epsilon_{\pi^{0}}}{\epsilon_{\gamma}} 
\end{equation} 
is then determined, 
where $N(D^+\gamma)/N(D^+\pi^{0})$ is the ratio of the number
of $D^{*+}$ decays observed in each mode, and 
$\epsilon_{\pi^{0}} / \epsilon_{\gamma}$ is
the relative efficiency for finding the $\pi^{0}$ or the $\gamma$ from the 
corresponding $D^{*+}$ decay.
Assuming that the three decay modes of the $D^{*+}$ add to unity
and defining 
$\Rp \equiv {\cal B}(D^{*+} \to D^0\pi^+)/{\cal B}(D^{*+} \to D^+\pi^0)$, 
one finds 
    $ {\cal B}(D^{*+}\rightarrow D^+\gamma)= 
           {\Rg}/{(\Rg+\Rp+1)} $,
    $ {\cal B}(D^{*+}\rightarrow D^+\pi^0)= 
           {1}/{(\Rg+\Rp+1)}$ and
    $ {\cal B}(D^{*+}\rightarrow D^0\pi^+)= 
           {\Rp}/{(\Rg+\Rp+1)}$.
Constraints on $\Rp$ can be obtained by combining the known phase space for 
$D^{*+}\rightarrow D^+\pi^0$ and $D^{*+}\rightarrow D^0\pi^+$ with
isospin conservation and the expected $p^3$ dependence of $p$-wave decay
widths to yield,
\begin{equation}
        \Rp=2 \left( \frac{p_{+0}}{p_{++}}\right)^3=2.199\pm0.064 
\label{Rp}
\end{equation} 
(where $p_{+0}$ and $p_{++}$ are the
momenta of the $D^0$ and $D^+$ in the $D^{*+}$ rest frame, respectively).
The theoretical uncertainty in this ratio is thought to be only of order 
1\%~\cite{Lepage}, so the error is dominated by those due to
the $M_{D^*}-M_D$ mass differences~\cite{PDG}.
This method has the advantage of avoiding large systematic uncertainties
due to the $D$ meson branching fractions and of canceling many systematic 
uncertainties associated with the $D^+$ reconstruction.

The analysis was performed using data accumulated by the CLEO~II
detector~\cite{CLEO} at the Cornell Electron Storage Ring (CESR).
The CLEO~II detector consists of three cylindrical drift chambers 
(immersed in a  1.5 T solenoidal magnetic field) surrounded by a 
time-of-flight system (TOF) and a CsI crystal electromagnetic (EM) 
calorimeter.  
The main drift chamber allows for charged particle identification 
via specific-ionization measurements ($dE/dx$) in addition to providing
an excellent momentum measurement.
The calorimeter is surrounded by a superconductor coil and an iron flux 
return, which is instrumented with muon counters.
 
A total of 4.7 fb$^{-1}$ of data were collected at 
center-of-mass energies on or near 
the $\Upsilon$(4S) resonance.
The Monte Carlo simulated events used to determine signal shapes and detection 
efficiencies were produced with a GEANT-based full detector simulation.  
Also, a continuum Monte Carlo sample 
(which contains roughly double the statistics of the data) was used
to test the analysis code and methods.  

Events were required to have three or more tracks and at least 15\% of
the center-of-mass energy deposited in the calorimeter. 
Each of the three tracks comprising a candidate 
$D^+ \to K^-\pi^+\pi^+$ decay was 
required to satisfy either the $K^-$ or $\pi^+$ hypothesis
at the $2.5\sigma$ level using $dE/dx$ alone, and then the triplet was 
required to satisfy the $K^-\pi^+\pi^+$ hypothesis, including TOF
information if available, with a $\chi^2$ probability greater than $10\%$.
The three tracks were then constrained to come from a common vertex, and the 
invariant mass of the triplet, under the $K^-\pi^+\pi^+$ 
hypothesis, was required to be within 10~MeV/$c^2$ ($\sim 1.5\sigma$) of the 
known $D^+$ mass.

Photon candidates were required to be in the best region of the calorimeter,
$|\rm{cos} \theta |<0.71$ (where $\theta$ is the polar angle between 
the EM cluster
centroid and the beam axis), with a cluster energy of at least 30~MeV.
It was further required that no charged particle track point 
within 8 cm of a crystal used in the EM cluster.
If the invariant mass formed by a pair of photons was within $2.5\sigma$ 
of the $\pi^{0}$ mass, taking into account the asymmetric $\pi^0$ line 
shape and the small momentum dependence of the mass resolution, 
the photons were identified as being from a $\pi^{0}$.  
The photons were then kinematically constrained to the $\pi^0$ mass to 
improve the $\pi^0$ momentum measurement. 

Photons from $D^{*+} \to D^+ \gamma$ decays were required to pass a  
lateral shower shape cut, which is 99\% efficient for isolated photons,
and not to form a $\pi^{0}$ when paired with any other photon.
The decay angle $\theta_\gamma$, defined as the angle of the $\gamma$
in the $D^{*+}$ rest frame with respect to the $D^{*+}$'s direction in the 
laboratory frame, was required to satisfy ${\rm cos} \theta_\gamma >-0.35$.
This cut helps to reduce the large combinatorial background that arises when
$D^{+}$ mesons are combined with soft photons moving in the opposite 
direction.

The combinatorial background was further reduced by requiring $x_{D^*}>0.7$,
where $x_{D^*}$ is the fraction of the maximum possible momentum
carried by the reconstructed $D^{*+}$.  This cut also removed any contribution
from $B\to D^*X$ events.  The cuts on ${\rm cos}\theta_{\gamma}$ and $x_{D^*}$
were determined to maximize $S^2/B$ ($S$ is signal and $B$ is background) 
by utilizing a large sample of $D^{*0} \to D^0 \gamma$ events from the data
as well as Monte Carlo simulated events.

The primary difficulty in this analysis is the small size of the signal, due
to the branching fraction, relative to a large combinatorial
background and, more importantly, relative to a background due to 
$D_s^{*+}$ radiative decays where $D_s^+ \to K^-K^+\pi^+$.  
Unlike the $D^{*+}$, the $D_s^{*+}$ almost always decays radiatively.
This is a major problem because the $M(D_s^{*+})-M(D^+_s)$ mass difference is 
$143.97\pm0.41$~MeV~\cite{Bartelt_PRL} and the $M(D^{*+})-M(D^{+})$ 
mass difference is 
$140.64\pm0.09$~MeV~\cite{PDG}, so these two processes cannot be separated 
in the mass difference plot because
the resolution in photon energy in the decay is $\sim 6$~MeV.
Misidentification of $D_s^+ \to K^-K^+\pi^+$ as $D^+ \to K^-\pi^+\pi^+$
can occur because the TOF and $dE/dx$ 
information used for particle identification does not adequately
separate $K$'s from $\pi$'s with momenta above $\sim 1$ GeV$/c$.
When reconstructed under the $K\pi\pi$ hypothesis, the two invariant mass 
distributions partially overlap, and any attempt to estimate
the fraction of $D_s^+$ under the $D^+$ peak will depend strongly on the 
resonant substructure of the $D_s^{+} \rightarrow K^-K^+\pi^+$ decay, as
well as the momentum distribution of the $D^{+}_s$'s.  
The large $D_s^+$  contribution to the lower $D^+$ sideband further 
complicates the analysis by preventing the use of this sideband in a
subtraction of combinatorial background.

\begin{figure}[t]
    \begin{center}
    \spwidth  
    \epsffile{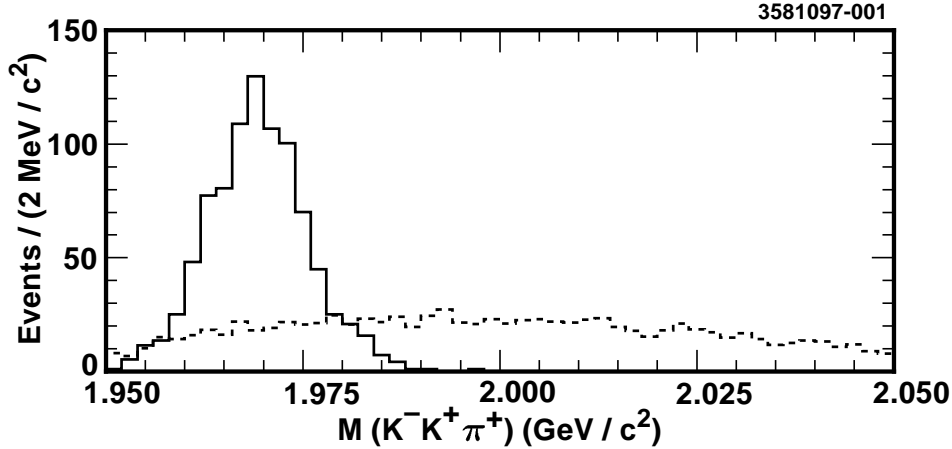}
    \vspace{0.75em}
    \caption{
        The $M(K^-K^+\pi^+)$ distributions for $D_s^{+}$ background (solid)
        and $D^{+}$ signal (dashed) Monte Carlo samples.}
    \vspace{-1.25em}
    \label{mds}
    \end{center}
\end{figure}
A means to veto $D^{*+}_s$ events,  independent of the decay's 
resonant substructure, is to require that the invariant mass of the three 
tracks reconstructed under the $K^-K^+\pi^+$ hypothesis be greater than a 
cut which removes all the $D_s^{*+}$ events.
An unwanted side effect of vetoing $D^{*+}_s$ events by this method is that
a cut in the $KK\pi$ mass distribution greatly distorts the $K\pi\pi$
mass distribution, making the relative normalization between the 
$D^+$ upper sideband and the signal region uncertain.  
Thus the use of a sideband subtraction to remove the combinatorial background
from the mass difference plot is impossible.
Fig.~\ref{mds} shows the Monte Carlo $K^-K^+\pi^+$ mass distribution found in 
$D_s^+$ decays and
that found in $D^+$ decays when one of the $\pi^+$'s is misidentified as
a $K^+$.
Since there are two possible tracks to assign the
$K^+$ mass, both combinations are tried, and the one yielding the smaller
mass is plotted.

\begin{figure}[t]
    \begin{center}
    \spwidth  
    \epsffile{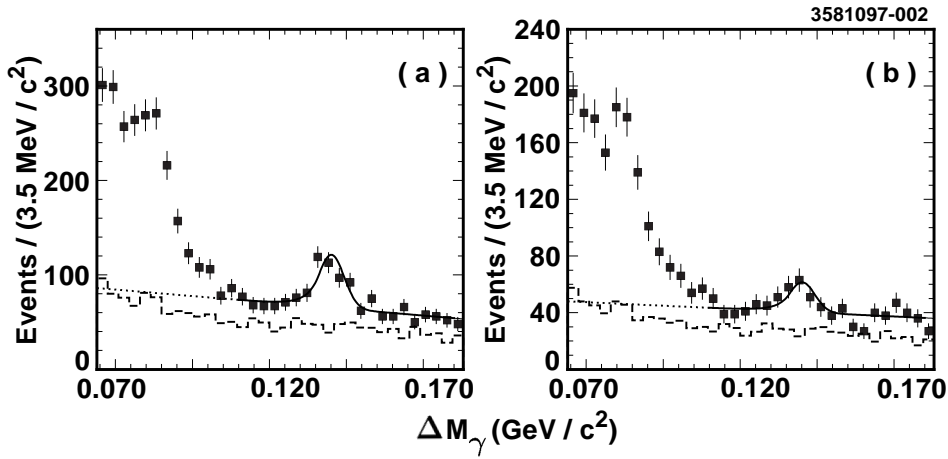}
    \vspace{0.75em}
    \caption{
	The $\Delta M_\gamma \equiv M(D^+ \gamma)-M(D^+)$ distributions for
        (a) data  before the $D_s^{*+}$ veto has been applied,
        (b) data after the tight $D_s^{*+}$ veto has been applied.
	   The large feature on the left of the plots is due to
	   $D^{*+} \to D^+ \pi^0$ where one of the photons from the
	   $\pi^0$ decay is not detected.  Monte Carlo studies 
	   indicate that this 
	   decay does not contribute to the signal region.
	   The dashed histograms are data taken from the upper $M(D^+)$
		sideband.}
    \vspace{-0.75em}
    \label{dg}
    \end{center}
\end{figure}
Fig.~\ref{dg}(a) shows the $\Delta M_\gamma$ distribution 
for events from the $M(D^+)$ signal region as well as for those 
from the $M(D^+)$ upper sideband (a region three times as wide as the
signal region starting  $\approx 3\sigma$ above the nominal $D^+$ mass). 
The $\Delta M_\gamma$ distribution 
for the combinatorial background found in the $M(D^+)$ sideband is quite 
flat under the signal region, justifying
the use of a first order polynomial in fitting this background.
No $D^{*+}_s$ veto has been applied to the data in 
Fig.~\ref{dg}(a), so a fair fraction of the events in this ``signal'' 
are $D_s^{*+}$ background.  The signal was fit with a modified Gaussian,
the parameters for which were obtained from
a large Monte Carlo sample of $D^{*+} \to D^+ \gamma$ events.
The systematic error in the fit parameters was estimated by studying 
data versus Monte Carlo differences in the very similar decay 
$D^{*0} \to D^0 \gamma$.
 
  Fig.~\ref{dg}(b) shows the $\Delta M_\gamma$ signal and sideband
distributions for events satisfying the $D_s^{*+}$ veto requirement that
$M(K^-K^+\pi^+)>1.990~{\rm GeV}/c^2$.
Monte Carlo indicates the fraction of $D_s^{*+}$ events passing this cut
is $0.002^{+0.003}_{-0.002}$, thus if the entire signal yield
($180\pm 26$ events) found in Fig.~\ref{dg}(a)
were due to $D_s^{*+}$ decays, $0.4^{+0.6}_{-0.4}$ events would be
expected in Fig.~\ref{dg}(b).
   The fit to the $\Delta M_\gamma$ distribution in Fig.~\ref{dg}(b) yields
   $68 \pm 19$ events.  When these data are refit with the signal constrained
   to be $0.4^{+0.6}_{-0.4}$ events, the $\chi^2$ of the fit increases by 15.8,
   corresponding to a significance of 4.0 standard deviations for the
   $D^{*+} \to D^+ \gamma$ signal.

\begin{figure}[t]
\begin{center}
    \spwidth  
    \epsffile{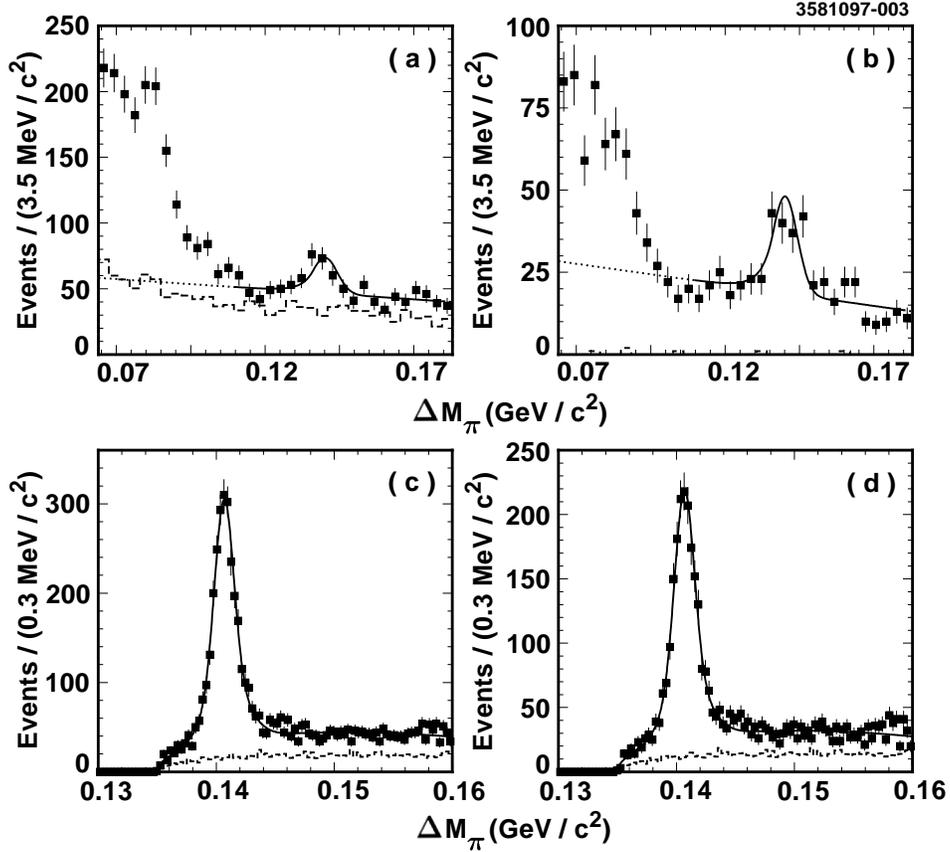}
    \vspace{0.75em}
    \caption{ 
        (a) $\Delta M_\gamma$ distribution for data after the ``optimal'' 
                $M(K^-K^+\pi^+)$ cut (the $D_s^{*+}$ veto) has been applied.
        (b) $\Delta M_\gamma$ distribution for the vetoed data.
        (c) $\Delta M_\pi$ distribution for data prior to the 
                $M(K^-K^+\pi^+)$ cut. 
        (d) $\Delta M_\pi$ distribution for data after the 
                $M(K^-K^+\pi^+)$ cut is applied.
	    The dashed histograms are data taken from the upper $M(D^+)$
		sideband.}
    \vspace{-0.75em}
    \label{final_prl}
\end{center}
\end{figure}

  The presence of $D^{*+} \to D^+ \gamma$ decays having been established,
the $D^{*+}_s$ veto was loosened to maximize $S^2/(S+B)$ as determined by the
Monte Carlo samples.
Fig.~\ref{final_prl}(a) shows the $\Delta M_\gamma$ distribution for the 
events which passed the optimized $D_s^{*+}$ veto.
The fraction of $D^+$ mesons passing the veto was determined 
by fitting the $\Delta M_\pi$ distribution before and after the veto was 
applied to the data.
This distribution was fit with a double Gaussian plus a background 
function~\cite{bkg} which simulates the expected threshold behavior.
Figs.~\ref{final_prl}(c) and \ref{final_prl}(d) 
show the $\Delta M_\pi$ distributions, 
along with the fits, used to determine the $D^{*+}_s$ veto efficiency 
for $D^+$ mesons.

The results of fitting the $\Delta M_\gamma$ distribution for events which
passed and for those which failed the $D_s^{*+}$ veto,
Figs.~\ref{final_prl}(a) and \ref{final_prl}(b) respectively, 
were: $N^{pass}_\gamma=87 \pm 21$, $N^{fail}_\gamma=95 \pm 16$ 
(statistical errors only).
Defining $N_+$ ($N_s$) as the total number of $D^{*+}$ ($D_s^{*+}$) in the
data, the branching ratio $\Rg$ was then extracted by solving the 
following pair of equations 
\begin{equation}
\begin{array}{ccccc}
   (1-\epsilon_+)N_+ &+&(1-\epsilon_s)N_s &=&  N^{fail}_\gamma\\
      \epsilon_+ N_+ &+&   \epsilon_s N_s &=&  N^{pass}_\gamma
\end{array}
\end{equation}
where $\epsilon_+$ is the fraction of $D^+$'s which pass the veto as 
determined by fitting the $\Delta M_\pi$ distributions 
($N_\pi^{pass}=1650 \pm 57$ and $N_\pi^{total}=2265 \pm 66$,
where the errors are statistical only),
and $\epsilon_s=0.037 \pm 0.007$ is the fraction of $D_s^{*+}$'s which escape 
the veto as determined by a Monte Carlo study.
We find 
\begin{equation}
\begin{array}{rcl}
\Rg &=&\frac{ N^{pass}_\gamma - \epsilon_s(N^{fail}_\gamma+N^{pass}_\gamma)} 
	   {  N_\pi^{pass} - \epsilon_s N_\pi^{total}} \times
      \frac{\epsilon_{\pi^{0}}}{\epsilon_{\gamma}} \\ 
    &=& 0.055 \pm 0.014 \pm 0.010
\end{array}
\end{equation}
where the ratio of efficiencies
$\epsilon_{\pi^{0}} / \epsilon_{\gamma}=1.066 \pm 0.064$.
From this branching ratio we can then extract the branching 
fractions shown in Table~\ref{branch}.
\begin{table}
\centering
\begin{minipage}[t]{4.5in}
\caption{The $D^*(2010)^\pm$ branching fractions determined from the 
measured ratio N($D^+\gamma$)/N($D^+\pi^0$).
The first uncertainty is statistical, the
second is experimental systematic and the third is that which arises from
the use of Eq.~(\ref{Rp}).
}
\begin{tabular}{c@{\hspace{.5in}}c@{\hspace{.5in}}c}
    Mode     & CLEO II  & PDG~\cite{PDG} \\ \hline  
$D^+ \gamma$ & ($1.68 \pm 0.42 \pm 0.29 \pm 0.03$)\%  
                                        & $(1.1^{+2.1}_{-0.7})$\% \\ 
$D^+ \pi^0 $ & ($30.73 \pm 0.13 \pm 0.09 \pm 0.61$)\%  
                                        & $(30.6 \pm 2.5)$\% \\ 
$D^0 \pi^+ $ & ($67.59 \pm 0.29 \pm 0.20 \pm 0.61$)\%
                                        & $(68.3 \pm 1.4)$\% \\
\end{tabular}
\label{branch}
\end{minipage}
\end{table}
The statistical uncertainty is dominated by the $D^+\gamma$ yields, and the
largest systematic uncertainty is due to variations in this yield
when the mean and width of the signal shape was varied by an amount suggested
by the $D^{*0} \to D^0\gamma$ data versus Monte Carlo comparison.  A similar
comparison was used to estimate the uncertainty introduced by 
the cos$\theta_\gamma$ cut.
Table~\ref{sys} lists the various sources of systematic uncertainty and gives
estimates for their impact on the measurement of $\Rg$.

In conclusion, we have observed, with $4\sigma$ significance, the radiative 
decay of the $D^{*+}$ and measured 
$ {\cal B}(D^{*+}\rightarrow D^+\gamma)/
  {\cal B}(D^{*+}\rightarrow D^+\pi^{0}) =  0.055 \pm 0.017 $
(statistical and systematic uncertainties added in quadrature).
Assuming Eq.~(\ref{Rp}) and that the three branching fractions of the
$D^{*+}$  add to unity, we find the results in Table~\ref{branch}. 
The hadronic branching fractions are in good agreement with the current PDG
averages, but with substantially reduced uncertainties (which are now 
dominated by the 3\% uncertainty in $\Rp$).
The $D^{*+}$ radiative branching fraction is in good agreement with 
theoretical 
expectations and the earlier upper limits set by CLEO~II~\cite{CLEOII} and 
ARGUS~\cite{ARGUS}.
The uncertainty in this branching fraction is due primarily to the large 
combinatorial background under the radiative signal, so one can 
expect that data taken with the new CLEO~II.5 detector, which includes 
a silicon tracker, to reduce this uncertainty further in the near future. 
\begin{table}
\centering
\begin{minipage}[t]{3.0in}
\caption{Estimates of the systematic uncertainties in the measurement of
$\Rg$.}
\begin{tabular}{c@{\hspace{.25in}}c}
efficiency ratio $\epsilon_{\pi^0}/\epsilon_{\gamma}$ & 6\%\\ 
fitting of background & 9\%  \\
fitting of signal & 13\%  \\ 
veto efficiency for $D^+_s$ (19\% on $\epsilon_s$)& 1\% \\
veto efficiency for $D^+$ (2\% on $\epsilon_+$)& 2\% \\
${\rm cos}\theta_\gamma>-0.35$ & 5\% \\ 
\end{tabular}
\label{sys}
\end{minipage}
\end{table}

We gratefully acknowledge the effort of the CESR staff in providing us with
excellent luminosity and running conditions.
This work was supported by 
the National Science Foundation,
the U.S. Department of Energy,
the Heisenberg Foundation,  
the Alexander von Humboldt Stiftung,
Research Corporation,
the Natural Sciences and Engineering Research Council of Canada, 
the A.P. Sloan Foundation, 
and the Swiss National Science Foundation.


\begin{thebibliography}{99}
\bibitem{CC}
Inclusion of the charge conjugate reaction is implied throughout this paper.
\bibitem{bkg}
The background function used is, 
$p_1 [ (\Delta M_\pi-M_{\pi^0})^{1/2} + p_2(\Delta M_\pi-M_{\pi^0})^{3/2})]$,
where $p_1$ and $p_2$ are parameters determined by the fit.
\end{thebibliography}
\end{document}